\documentclass[prb,twocolumn,aps,floats,amssymb,amsmath,floatfix]{revtex4}
\usepackage{epsf}
\def\bea{\begin{eqnarray}}
\def\eea{\end{eqnarray}}

\def\beq{\begin{equation}}
\def\eeq{\end{equation}}
\def\e{{\rm e}}
\def\squareforqed{\hbox{\rlap{$\sqcap$}$\sqcup$}}
\def\qed{\ifmmode\squareforqed\else{\unskip\nobreak\hfil
\penalty50\hskip1em\null\nobreak\hfil\squareforqed
\parfillskip=0pt\finalhyphendemerits=0\endgraf}\fi}

\def\Im{{\rm Im\,}}

\def\<{\langle}
\def\>{\rangle}

\newcommand{\ket}[1]{\ensuremath{\vert{#1}\rangle}}
\newcommand{\bra}[1]{\ensuremath{\langle{#1}\vert}}

\newcommand{\bracket}[2]{\ensuremath{\langle{#1}\mid{#2}\rangle}}

\newcommand{\ovs}{s_{\rm max}}

\begin{document}

\title{Charge-density-wave instabilities driven by multiple umklapp
  scattering} 

\author{Peter Schmitteckert and Ralph Werner}
\affiliation{Institut f\"ur Theorie der
Kondensierten Materie, Universit\"at Karlsruhe, 76128 Karlsruhe, Germany}

\date{\today}


\begin{abstract}
We show that the concept of umklapp-scattering driven instabilities in
one-dimensional systems can be generalized to arbitrary multiple
umklapp-scattering processes at commensurate fillings given that the
system has sufficiently longer range interactions. To this end we
study the fundamental model system, namely interacting spinless
fermions on a one-dimen\-sional lattice, via a density matrix
renormalization group approach. The instabilities are investigated via
a new method allowing to calculate the ground-state charge stiffness
numerically exactly. The method can be used to determine other
ground state susceptibilities in general. 
\end{abstract}
\maketitle
%


\section{Introduction}

The relevance of umklapp-scattering terms in
one-dimensional systems with commensurate filling has long been
established.\cite{Soly79} Besides the general theoretical interest in
one-dimensional systems because of the existing exact solutions to
some of them \cite{KM97} and the field-theoretical description of the
low energy physics,\cite{GNT98} the application of one-dimensional
models to the rich phase diagrams of quasi one-dimensional systems
such as the Bechgaard and Fabre salts \cite{Bour00} has a long
standing history. The degeneracy of the ground state leads to the
presence of solitonic excitations in the one-dimensional systems and
recent studies focus on the question to what extent these collective
modes are observable \cite{ET02} in the systems under investigation.

Here we focus on the generic instabilities of systems of fermions
without the internal magnetic degree of freedom. The Hamiltonian under
consideration is   
\begin{equation}\label{Hgeneric}
H = -t \sum_{l=1}^M \left(c^\dagger_{l+1} c^{\phantom{\dagger}}_{l} + {\rm
    h. c.} \right)
    + \sum_{s=1}^{\ovs} V_s \sum_{l} n_{l} n_{l+s}\,,
\end{equation}
with fermionic creation ($c^\dagger_{l}$) and annihilation
($c^{\phantom{\dagger}}_{l}$) operators at site $l$, nearest-neighbor
hopping amplitude $t$, density-density interaction amplitude $V_s$
between fermions on sites of separation $s$ and density operators $n_{l} =
c^\dagger_{l} c^{\phantom{\dagger}}_{l}$. Energies are given in units
of $t$. The decay of the interaction with distance requires in general
$V_{s+1} < V_s$, with the exception of zig-zag chains and some spin
systems where the nearest-neighbor interaction is suppressed by the
lattice geometry. We apply periodic boundary conditions, i.e.,
$c_{M+1} \equiv c_{1}$.\cite{Antiquote} The model in Eq.\
(\ref{Hgeneric}) is directly applicable to spin 1/2 chains that can be
mapped onto chains of spinless fermions.\cite{ETD97} Furthermore, its
widely studied continuum representation \cite{Hald82b} is similar to
the charge sector of the bosonized Hubbard model.\cite{ET02,YTS01}
Finally, Eq.\ (\ref{Hgeneric}) allows to study the interplay of
ordered phases with different modulation wave vectors, which proves
useful for the understanding of materials that exhibit multiple phase
transitions.\cite{RP99}

The purpose of this paper is threefold.
(i) We present a novel method to determine the ground-state curvature or,
equivalently, the ground-state charge stiffness.
(ii) The approach allows for an accurate identification of
charge-density-wave (CDW) instabilities at commensurate fillings. The
phase diagrams for various relevant sets of parameters are shown. 
(iii) We discuss the physical insight that can be gained from the
continuum representation of Eq.\ (\ref{Hgeneric}) and by comparison
with the numerical results show the importance of lattice effects.

To this end we present a detailed density matrix renormalization group (DMRG)
study \cite{DMRG} of the lattice model Eq.\
(\ref{Hgeneric}). Sufficiently large system sizes are accessible to
model the thermodynamic limit without the shortfall of neglecting
possible lattice effects.


\section{Numerical approach}

The ground state curvature $C$ is defined as 
\begin{equation}\label{defC}
    C = M \left. \frac{ \partial^2 E_0(\phi) }
      {\partial\phi^2}\right|_{\phi=0},
\end{equation}
where $\phi$ is a magnetic flux that penetrates our system which is
closed to a ring. The magnetic flux inside the ring can be gauged into
a single bond leading to twisted boundary conditions 
$ c_{1} = \e^{\imath \phi} c_{M+1} $. 
According to Kohn \cite{Kohn64} $C$ is equivalent to the Drude weight
of the $T=0$ dc conductivity. The curvature in Eq.\ (\ref{defC}) is
the second order coefficient of the Taylor series of $E_0(\phi)$,
i.e., $E_0(\phi) = E_0(0) + E^{(1)} \phi + \frac{C}{2M} \phi^2 +
O(\phi^3)$. In order to compute the coefficient exactly we expand  
$H(\phi)$ and the ground-state wave function $\ket{\Psi_0(\phi)}$ at
$\phi=0$ in second order. $H$ is given via    
\begin{equation}\label{Hexpand}
    H(\phi) = H^{(0)} + \imath\phi J - \frac{\phi^2}{2} T +
    {\rm O}(\phi^3)\,,\\
\end{equation}
where $H^{(1)} = \imath J$ is the current operator with
\begin{equation}
\label{defJ}
        J = c^\dagger_{M} c^{\phantom{\dagger}}_{1} -
        c^\dagger_{1} c^{\phantom{\dagger}}_{M} 
\end{equation}
and 
\begin{equation}
  \label{defT}
  T = H^{(2)} = - (c^\dagger_{M} c^{\phantom{\dagger}}_{1} +
  c^\dagger_{1} c^{\phantom{\dagger}}_{M})
\end{equation}
is the kinetic energy on the bond between the $M$th and first site.
Due to the normalization of the wave function
$
    \bracket{\Psi(\phi)}{\Psi(\phi)} = 1 =
    \bracket{\Psi^{(i)}}{\Psi^{(i)}}
$
the first order correction $\ket{\Psi^{(1)}}$ in the expansion
$\ket{\Psi(\phi)} = \ket{\Psi^{(0)}} + \phi \ket{\Psi^{(1)}} +
{\rm O}(\phi^2)$ is orthogonal to the ground state $\ket{\Psi^{(0)}}$
at zero flux and is obtained by solving the set of linear equations
\begin{equation} \label{LGS_Psi1}
    \left( H^{(0)}-E^{(0)}_0 \right) \ket{\Psi^{(1)}} =
    \left(E^{(1)}-\imath J\right) \ket{\Psi^{(0)}}\, .
\end{equation}
on the subspace that is orthogonal to the ground state. The resulting
curvature is given by 
\begin{equation}
    C = M\bra{\Psi^{(0)}} T
    \ket{\Psi^{(0)}}  + 2 M \bra{\Psi^{(0)}} \imath J - E^{(1)}
    \ket{\Psi^{(1)}} 
\end{equation}
with $E^{(1)} = \bra{\Psi^{(0)}} H^{(1)} \ket{\Psi^{(0)}}$.

It must be stressed that this approach is exact. Therefore the
accuracy of our numerical results is only limited by the truncation
error of the DMRG. Consequently any ambiguity in the determination of
$C$ incurred through a numerical finite difference approximation of
the derivative of $E_0(\phi)$ is avoided. Moreover, our procedure
allows to calculate $C$ without referring to a system with a finite
flux that could destroy the commensurability effects we want to
measure. 

For systems with time-reversal symmetry and a nondegenerate ground
state one finds $E^{(1)}= 0$ and from Eq.\ (\ref{LGS_Psi1}) follows  that
$\ket{\Psi^{(1)}}$ is purely imaginary. In turn it follows that $C$
can be calculated by using real numbers only. Therefore solving the
linear set of equations (\ref{LGS_Psi1}) and storing 
the additional target vectors $\Im\ket{\Psi^{(1)})}$ and
$J\ket{\Psi^{(1)})}$ are outweighed by the fact that we can restrict
ourselves to real numbers. Since $\bracket{\Psi^{(1)}}{\Psi^{(0)}} =
0$, solving Eq.\ (\ref{LGS_Psi1}) is stable and can be performed by a
mixture of standard preconditioned minimal residue and conjugate
gradient methods which we extended by a projection on the space
orthogonal to the subspace of the ground state(s). The extension to
degenerate ground states is straightforward and can be performed
similar to Brillouin-Wigner perturbation theory.\cite{Zima69} 
For the small systems ($M<30$) we keep 300 to 400 states per block $A$
and $B$, while we keep up to 1000 states per block for the larger
systems. In addition to the ground state $\ket{\Psi^{(0)}}$ and the
auxiliary states $J \ket{\Psi^{(0)}}$ and $\ket{\Psi^{(1)}}$ we target
at least for the first four excited eigenstates to treat degeneracies
properly.


\section{Phase diagrams}

\begin{figure}
   \epsfxsize=0.5\textwidth
   \centerline{\epsffile{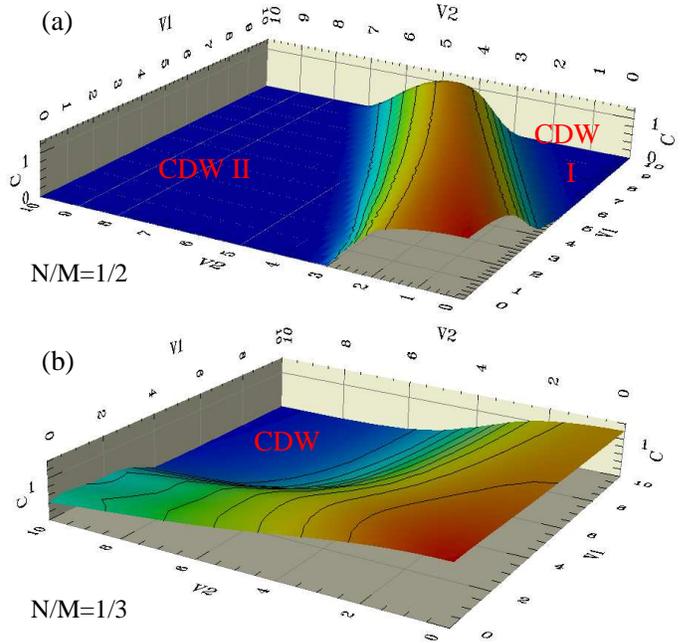}}
   \caption{\label{V1V2phases} Curvature of the (a) half and (b)
   one-third-filled chain with (a) $M=32$ and (b) $M=24$ sites as a
   function of the nearest- ($V_1$) and next-neares-neighbor ($V_2$)
   interaction strength. Charge order implies $C = 0$. The ordering
   pattern are discussed in the text. In (a) the CDW phase I
   doubles the unit cell,  phase II quadruples the unit cell and the
   ground states are twofold and fourfold degenerate, respectively. In
   (b) the ordered phase triples the unit cell with triply degenerate
   ground state.}
   \end{figure}

In order to investigate the general features
the model in Eq.\ (\ref{Hgeneric}) we study its phase diagram as a
function of the nearest- ($V_1$) and next-nearest-neighbor ($V_2$)
interaction. The result is presented in Fig.\ \ref{V1V2phases}, where
the curvature is plotted versus $V_1$ and $V_2$. Charge order implies
$C(V_1,V_2) = 0$. In Fig.\ \ref{V1V2phases}(a) the system is shown at
half filling for a chain of length 32. In CDW phase I the ground state
is twofold degenerate \cite{Finitequote} with ordering pattern  
($\bullet$$\circ$$\bullet$$\circ$) and
($\circ$$\bullet$$\circ$$\bullet$). Here $\circ$ denotes a vacant and
$\bullet$ denotes an occupied site. In phase II the ground state is
fourfold degenerate with ordering pattern
($\bullet$$\bullet$$\circ$$\circ$),
($\circ$$\bullet$$\bullet$$\circ$),
($\circ$$\circ$$\bullet$$\bullet$), and
($\bullet$$\circ$$\circ$$\bullet$).
The transition between phases I and II along the line $V_1=2V_2$ is
expected from simple potential energy arguments for the corresponding
real-space ground-state configurations. The finite width of the
uniform region is due to a gain in kinetic energy near the line
$V_1=2V_2$, where the charge ordering effects are reduced by the
competition of the two potential energies. Figure \ref{V1V2phases}(b)
shows the curvature of a chain of length 24 at $1/3$ filling. The
ordered phase is observed for sufficiently large values of
$
V_1,V_2\ \raisebox{-0.7ex}{$\stackrel{\mbox{$>$}}{\sim}$}\ 4
$
and has a triply degenerate ground state  with ordering pattern
($\bullet$$\circ$$\circ$$\bullet$$\circ$$\circ$),
($\circ$$\bullet$$\circ$$\circ$$\bullet$$\circ$), and
($\circ$$\circ$$\bullet$$\circ$$\circ$$\bullet$).

\begin{figure}
   \epsfxsize=0.48\textwidth
   \centerline{\epsffile{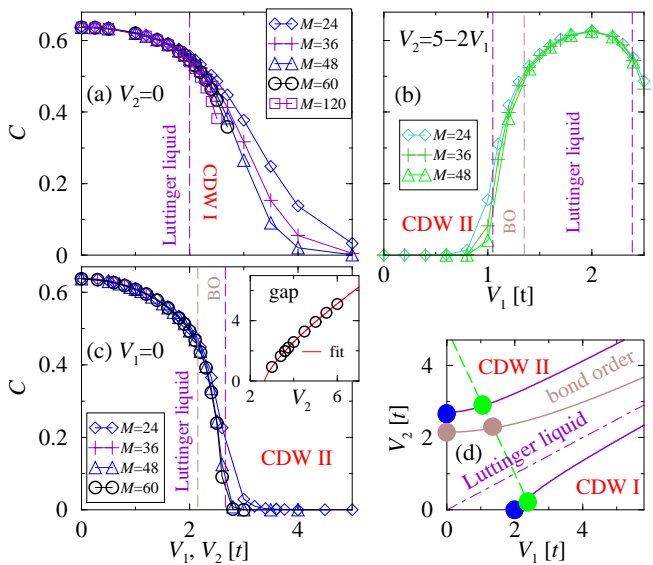}}
   \caption{\label{cuts} Curvature at half filling as a
   function of the interaction parameters for different system
   sizes along cuts through Fig.\ \protect\ref{V1V2phases}(a). Panel
   (a) $V_2=0$. Panel (b) $V_2 = 5 - 2 V_1$. Panel (c) $V_1=0$, the
   inset shows that the gap in CDW~II opens as $\Delta_{\rm II} \sim
   (V_2 - V_{2,\rm c})^{3/4}$. Panel (d) shows a sketch of the
   resulting phase diagram. The dashed line indicates the direction of
   the cut shown in panel (b), the dash-dotted lines shows $2 V_2 =
   V_1$. Dots mark the transition points determined herein.}
   \end{figure}

Figure \ref{cuts} shows the finite size effects of the curvature at
half filling via cuts through Fig.\ \ref{V1V2phases}(a) for various
system sizes. Figure \ref{cuts}(a) shows $C(V_1,0)$, panel (b)
$C(0,V_2)$, and panel (c) $C(V_1,10-2V_1)$. For $V_2=0$ and $M\to
\infty$ it is known from Bethe anstatz calculations\cite{dCG66} that
in the ordered phase for $V_1 > V_{1,\rm c} = 2$ an excitation gap
opens exponentially\cite{YY66c} as $\sim \exp\left( -\pi^2 /
  \sqrt{8(V_{1}-V_{1,\rm c})}\right)$. As a result the convergence of
the curvature to the thermodynamic limit in CDW~I near the transition
point is very slow [Fig.\ \ref{cuts}(a) and Ref.\
\onlinecite{LCS01}]. In contrast we find a gap in CDW~II obeying the
power law $\Delta_{\rm II}\big|_{V_1=0} \sim (V_2 - V_{2,\rm
  c})^{3/4}$ as shown in the inset of Fig.\ \ref{cuts}(c). The reason
for the different behavior lies in the bond-ordered phase (BO)
equivalent to that found in frustrated Heisenberg chains,\cite{SA01}
which we find strong evidence for between the Luttinger liquid (LL)
and CDW~II phases, see Fig.\ \ref{cuts}(d).

\begin{figure}
   \epsfxsize=0.41\textwidth
   \centerline{\epsffile{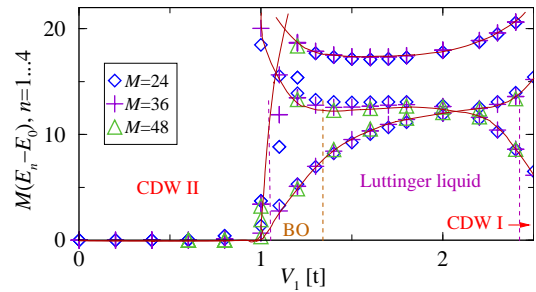}}
   \caption{\label{D5aFig} Level crossings at the phase boundaries
   between the BO and CDW II phase and changes of sign of the finite
   size effects of $M(E_1-E_0)$ at the boundaries of the LL phase
   along the line $V_2 = 5 - 2 V_1$. System sizes are $M=24$
   (diamonds), $M=36$ (crosses) and $M=48$ (triangles). Lines are
   guides to the eye.}   
   \end{figure}

The numerical determination of the critical values of the interaction
is difficult because of the smallness of the gap near the transitions.
It turns out though that the finite size effects of $M(E_1-E_0)$
change sign at the phase transition between the LL and the ordered BO
and CDW~I phases, i.e., $\lim_{M\to\infty} \partial M(E_1-E_0) /
(\partial M)|_{V_1=V_{1,c}} = 0$. On the other hand, in the CDW~II
phase systems with sizes $M{\rm mod}4 = 0$ have much smaller finite
size effects than for $M{\rm mod}4 = 2$ while in the BO phase both are
equivalent. Moreover, the systems exhibit well defined level crossings
between the bond ordered and the CDW phases.\cite{SA01} The effects
are illustrated in Fig.\ \ref{D5aFig} for $M=24$ (diamonds), $M=36$
(crosses), and $M=48$ (triangles) for a cut along the line $V_2 = 5 - 2
V_1$ for the four lowest excited states. Lines are guides to the
eye. Note that for the actual parameter determination systems of sizes
up to 60 sites where studied near the phase transitions.

Combining different approaches we estimate for $V_1=0$ a value of
$V_{2,\rm c}=2.66\pm0.1$ while along the line $V_2 = 5 - 2V_1$ the
critical points into the CDW phases are $(V_{1,\rm c},V_{2,\rm c}) \in
\{(1.05\pm0.05,2.9\pm0.1);(2.4\pm0.05,0.2\pm0.1)\}$. The transitions
from the LL to the BO phase occur at $(V_{1,\rm b},V_{2,\rm b}) =
(0,2.15\pm0.1)$ and $(1.35\pm0.05,2.3\pm0.1)$. The resulting phase
diagram is sketched schematically in Fig.\ \ref{cuts}(d). Thick dots
are transition values determined numerically within this work. 
The numerical analysis for large values of $V^*=20,50,100$ with $V_2
= V^* - 2 V_1$ suggests that for $V^* \gg t$ the uniform region (BO
and LL phases) between CDW~I and CDW~II extends along the line
$V_1=2V_2$ with a width of $\sim \sqrt{5}\,t$.\cite{3phasesquote}
At sufficiently high interaction strength the LL phase disappears. The
triple point between the BO, LL, and CDW~I phase can be estimated to
be near $(V_{1,\rm t},V_{2,\rm t}) \approx (8.2,3.6)$, see also 
Sec.~\ref{sectionCR}. 


\section{Commensurability effects}

\begin{figure}
   \epsfxsize=0.40\textwidth
   \centerline{\epsffile{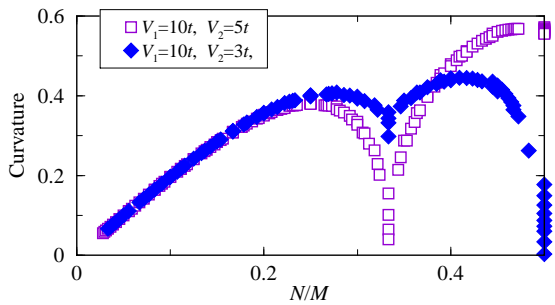}}
   \caption{\label{g_10_10}
   Curvature as a function of filling for different numbers of
   fermions $2 \le N \le M/2$ at system sizes $18 \le M \le
   60$. Diamonds: $V_1=10$ and $V_2=3$ with charge order at $N/M=1/2$
   and $N/M=1/3$. Squares: $V_1=10$ and $V_2=5$ with charge order only
   at $N/M=1/3$ (c.f.\ Fig.\ \protect\ref{V1V2phases}).}
   \end{figure}

Any deviation from commensurate
filling in one-dimensional systems leads to the presence of solitonic 
excitations even at zero temperature which disorder the system due to
the degeneracy of the ground state.\cite{ET02} Our approach allows to
demonstrate the sensitivity of our one-dimensional model system to
the commensurability of the filling very nicely as is shown in Fig.\
\ref{g_10_10}. The diamonds show the curvature for different fillings
at different system sizes $18 \le M \le 60$ for $V_1=10$ and $V_2=3$,
i.e., the system is charge ordered both for fillings of $1/2$ and
$1/3$ as can be seen from Figs.\ \ref{V1V2phases}(a) and
\ref{V1V2phases}(b). Since the values of interaction are such that the
system at $N/M=1/3$ is close to the transition to the non-ordered
phase the convergence is slow at that point. The squares show the
curvature for $V_1=10$ and $V_2=5$, where no charge ordering is
observed at half filling due to the competition between $V_1$ and
$V_2$. Note that the results are symmetric about $N/M=1/2$ because of
particle-hole symmetry.

\begin{figure}
   \epsfxsize=0.40\textwidth
   \centerline{\epsffile{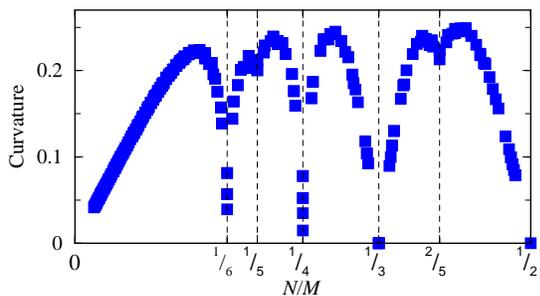}}
   \caption{\label{g_120_60}
   Curvature for $\ovs = 5$ as a function of filling for different
   numbers of fermions $2 \le N \le M/2$ at system sizes $20 \le M \le
   48$ with $V_1=120$, $V_2=60$, $V_3=30$, $V_4=15$, and $V_5=6$ with
   charge order instabilities at $N/M=1/6$, $N/M=1/5$, $N/M=1/4$,
   $N/M=1/3$, and $N/M=1/2$. Note that for the strong interactions
   under investigation there is also a second-harmonic 
   anomaly at $N/M=2/5$. The result supports the notion that
   instabilities occur only at fillings with $N/M \ge 1/(\ovs + 1)$.}
   \end{figure}

The results for $V_1,V_2\neq 0$ discussed in
detail above can readily be generalized to systems with longer range
interaction. For example, for $V_1,V_2,V_3\neq 0$ and sufficiently
large $V_3$ we find a phase with ordering pattern
($\bullet$$\circ$$\circ$$\circ$) and four-fold-degenerate ground
state at quarter filling. Since a detailed discussion would be
a simple extension of the case studied above it is omitted here. The
crucial result that follows from the generalization is that for all
systems under investigation ($\ovs=1,2,3,4,5$) an instability
at commensurate filling $N/M$ is only observed for sufficiently
long-ranged interaction, i.e., $V_s \neq 0\ \forall\ s \le \ovs$ with
$\ovs = M/N-1$. The case of $\ovs = 5$ is illustrated in Fig.\
\ref{g_120_60} for $V_1=120$, $V_2=60$, $V_3=30$, $V_4=15$, and
$V_5=6$ and exhibits charge density instabilities only for $N/M \ge
1/6$. Note that for sufficiently strong interaction there are
also second-harmonic anomalies such as the one at $N/M=2/5$ in Fig.\
\ref{g_120_60}.

The notion of instabilities only for sufficiently long ranged
interaction is quite intuitive when considering the system under
investigation in real space, because the system cannot order when the
mean distance between particles is larger than the interaction range.


\section{Continuum representation}\label{sectionCR}

\begin{figure}
   \epsfxsize=0.40\textwidth
   \centerline{\epsffile{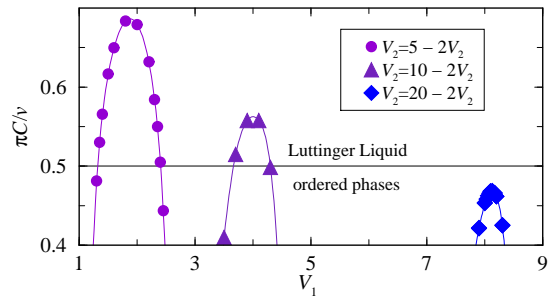}}
   \caption{\label{KofV}
   Numerical estimate $\pi C/v$ for the LL parameter $K$ for cuts
   along $V_2 = 5 - 2V_1$ (circles), $V_2 = 10 - 2V_1$ (triangles),
   and $V_2 = 20 - 2V_1$ (diamonds). For $0.5 < \pi C/v = K$ the
   system is in the LL regime, for $0.5 > \pi C/v$ in an ordered
   phase. For $V_1\sim 8$ the system is close to the tricritical point
   where the LL phase disappears for larger interaction strength. The
   numerical precision of the method to determine $\pi C/v$ can be
   estimated at $V_1=2$ and $V_2=0$ and amounts to about $\pm 5$\%.}   
   \end{figure}

The low energy properties of the 
Hamiltonian Eq.\ (\ref{Hgeneric}) can be approximated by its
field-theoretical equivalent obtained by the standard Abelian
bosonization technique.\cite{GNT98,Hald82b} The kinetic and 
forward-scattering parts lead to the free Hamiltonian 
\begin{equation}\label{H0bose}
{\cal H}_0 =  \frac{v}{16\pi} \int_{-\infty}^\infty d x
\left(K\ \Pi^2  + \frac{1}{K} \left(\partial_x\,
         \phi \right)^2\right)
\end{equation}
with excitation velocity $v$, Luttinger liquid parameter $K$, Bose
fields $\phi(x)$, and their conjugate momenta $\Pi(x)$. The continuum
coordinate is $x = \lim_{a\to 0} l a$ with the lattice constant $a$
and $M\to \infty$. The umklapp-scattering part is
\begin{equation}\label{Hubose}
{\cal H}_{\rm u} =  \lim_{a\to 0} \sum_s 
     \frac{V_s}{2(\pi a)^2} \int_{-\infty}^\infty d x
\cos\left[\sqrt{2} \phi - (4k_{\rm F}-G) x \right]
\end{equation}
with Fermi wave vector $k_{\rm F} = \pi \rho/a$ where $\rho = N/M \le
1$ is the number of fermions per site. $G$ is a reciprocal lattice
vector. For sufficiently small $K < 1/2$ and if $4k_{\rm F} = 2\pi/a$,
Eq.\ (\ref{Hubose}) yields a relevant contribution to the total
Hamiltonian ${\cal H} = {\cal H}_{0} + {\cal H}_{\rm u}$ which then
represents a sine-Gordon model.\cite{YTS01} Indeed, using the
finite-size scaling of the ground-state energy per site in the LL  
phase \cite{KBI93,SA01} $\frac{E_0(M)}{M} = e(\infty) -
\frac{\pi}{6M^2} v$, where $v$ is the velocity of the elementary
excitations, and the relation $C=\frac{v}{\pi} K$ we find $K > 1/2$
in the LL phase, $K \approx 1/2$ at the LL-phase boundary, and $K <
1/2$ elsewhere within the numerical precision of the method\cite{SA01}
of about $\pm 5$\%. The latter can be estimated at $V_1=2$ and
$V_2=0$. This is illustrated in Fig.\ \ref{KofV} for cuts along $V_2 =
5 - 2V_1$ (circles), $V_2 = 10 - 2V_1$ (triangles), and $V_2 = 20 -
2V_1$ (diamonds) and confirms the applicability of Eqs.\
(\ref{H0bose}) and (\ref{Hubose}) for $\rho=1/2$.

The obvious shortfall of Eq.\ (\ref{Hubose}) is that the operator in
the integrand does not depend on the range of the interaction $s$ as a
direct consequence of the continuum limit. The numerical study herein
proves that longer range interactions lead to instabilities also for
commensurate fillings other than $\rho = 1/2$ or, equivalently, for
$k_{\rm F} \neq G/4$, if the interaction range is larger than or equal
to the mean distance between the fermions on the lattice, i.e., for
$(\ovs + 1) \ge 1/\rho$. In order to reproduce this result within the
continuum model, Eq.\ (\ref{Hubose}) must be modified to
\begin{equation}\label{tHubose} \tilde{\cal H}_{\rm u} =
  \lim_{\alpha\to 0} \sum_{r=1}^\infty \frac{A_r}{2(\pi \alpha)^2}
  \int_{-\infty}^\infty d x 
  \cos\!\left[\sqrt{2} \kappa_r \phi - k_{\rm eff} x \right] 
\end{equation}
with $k_{\rm eff} = 2(r+1)k_{\rm F} - mG$. The presence of terms of
the form of Eq.\ (\ref{tHubose}) with $A_r\neq 0\ \forall\ r$ follows
from the perturbative inclusion of lattice corrections which yields
$\kappa_r = (r + 1)/2$.\cite{RA00} 
At any commensurate filling, where $2(r+1)k_{\rm F} = 2\pi
m/a$, the term for $r=m/\rho-1$ is relevant if $K < 0.5
\kappa_r^{-2}$.\cite{GNT98,ET02}

Preliminary numerical results suggest that $K$ indeed is decreased
with increasing interaction range. For example, near the phase
transition in the case of 1/3 filling, where $\ovs=2$, we find
$K \sim 2/9$ indicating that the term with $\kappa_2=3/2$ indeed
becomes relevant as expected. This leads to the conjecture that the
breakdown of the LL phase is determined by a critical value of
$K=2/(\ovs+1)^2$. Further support for this result requires a detailed
analysis of the dependence of the value of $K$ at the LL phase
boundaries as a function of $\ovs$ which is numerically involved and
beyond the scope of the present study. 

The excitations described by the interaction term $\tilde{\cal H}_{\rm
u}$ are $r+1$ electrons that are scattered from the left Fermi point
to the right Fermi point and vice versa. It can be referred to as
multiple ($r+1$) umklapp scattering. At incommensurate fillings the
integrand in Eq.\ (\ref{tHubose}) is oscillatory, the umklapp term is
effectively averaged out, and no instability is observed.



\section{Conclusions}

We have shown that (i) the exact zero-flux
determination of the ground-state curvature (or charge stiffness) is a 
powerful tool to determine numerically instabilities in interacting 
one-dimensional Fermion systems. Note that the method can easily be
adapted to determine ground state susceptibilities in general. (ii) The
application to systems of spinless fermions with sufficiently long-range
density-density interaction $V_s$, i.e., for $\ovs \ge
M/N-1$, leads to a comprehensive understanding of the rich phase
diagrams at commensurate fillings. The approach impressively
demonstrates the breakdown of long-range order for any incommensurate
filling. (iii) Lattice effects need to be included properly in the
continuum model in order to reproduce the numerically observed
instabilities. Within the adapted field-theoretical approach they find
an interpretation as driven by multiple ($mM/N$) umklapp-scattering
processes.



\begin{acknowledgments}

We thank A.\ Rosch for
discussions and for drawing our attention to the bond-order
instability. We thank M.\ Vojta and P.\ W\"olfle for instructive
discussions and J.\ Walter for boost::ublas. The work was supported by
the Center for Functional Nano\-struc\-tures of the Deutsche
Forschungsgemeinschaft within project B2.  

\end{acknowledgments}


\begin{thebibliography}{10}
\bibitem{Antiquote}
To be specific, we use (anti-) periodic boundary conditions for (even) odd $N$
  to avoid odd-even effects.
\bibitem{Finitequote}
The finite size scaling of a $m$-fold degenerate ground state is such that in a
  system with $N$ fermions the ground state is separated from a ($m-1$)-fold
  degenerate multiplet with a finite size gap that scales as $\Delta_{\rm fs}
  \sim N t^N V_s^{1-N}$.
\end{thebibliography}

\end{document}